\documentclass[%
 reprint,
 superscriptaddress,
 amsmath,amssymb,
 aps,
 apl,
]{revtex4-2}

\usepackage{float}
\usepackage{subcaption}
\restylefloat{figure}
\usepackage{graphicx}
\usepackage{dcolumn}
\usepackage{bm}
\usepackage{inputenc}
\usepackage{multirow}
\usepackage{upgreek} 
\usepackage{hyperref}

\hypersetup{
colorlinks=true,
linkcolor=blue,
citecolor=blue,
urlcolor=blue
}

\usepackage{listings}

\graphicspath{{./images/}}

\usepackage{etoolbox}

\makeatletter
\patchcmd{\frontmatter@abstract@produce}
  {\vskip200\p@\@plus1fil
   \penalty-200\relax
   \vskip-200\p@\@plus-1fil}
  {}
  {}
  {}
\makeatother

\begin{document}

\title{\texttt{Xsorb}: a software for identifying the most stable adsorption configuration and energy of a molecule on a crystal surface}

\author{Enrico Pedretti}
    \affiliation{Department of Physics and Astronomy, University of Bologna, 40127 Bologna, Italy}
\author{Paolo Restuccia}
    \affiliation{Department of Physics and Astronomy, University of Bologna, 40127 Bologna, Italy}
\author{M. Clelia Righi}
    \email[]{clelia.righi@unibo.it}
    \affiliation{Department of Physics and Astronomy, University of Bologna, 40127 Bologna, Italy}

\begin{abstract}
Molecular adsorption is the first important step of many surface-mediated chemical processes, from catalysis to  tribology. This phenomenon is controlled by physical/chemical interactions, which can be accurately described by first principles calculations. In recent years, several computational tools have been developed to study molecular adsorption based on high throughput/automatized approaches. However, these tools can sometimes be over-sophisticated for non-expert users. In this work, we present \texttt{Xsorb}, a Python-based code that automatically generates adsorption configurations, guides the user in the identification the most relevant ones, which are then fully optimized. The code relies on well-established Python libraries, and on an open source package for density functional theory calculations. We show the program capabilities through an example consisting of a hydrocarbon molecule, 1-hexene, adsorbed over the (110) surface of iron. The presented computational tool will help users, even non-expert, to easily identify the most stable adsorption configuration of complex molecules on substrates and obtain accurate adsorption geometries and energies.
\end{abstract}


\maketitle


\section{Introduction}
\label{sec:intro}

Molecular adsorption, i.e., the binding of a molecule over a surface, is a necessary prerequisite to any surface-mediated chemical process, from catalysis~\cite{Bligaard-2004,Andersson-2006} to molecular electronics~\cite{Joachim-2005}, biomedicine~\cite{Kasemo-2002}, electrochemistry~\cite{Hinnemann-2005}, friction~\cite{Neville2007,Peeters2019} and corrosion~\cite{Finsgar2014,Kousar2021}. For example, friction is ubiquitous whenever moving components are in contact. It causes significant energy losses and undermines the functionality of devices, ultimately leading to their failure. A strategy to reduce friction and wear in engines is based on the use of lubricant additives, i.e., molecules added in base oils that first adsorb over the sliding substrate and then react with it by forming protective, lubricious films~\cite{Minami-2017}. Similarly, corrosion damages many materials comprising carbon steel, which has a low intrinsic resistance to corrosive processes~\cite{Kermani-2003}. It is, therefore, crucial to use molecular corrosion inhibitors that create a protective coating, making the surface inert, more resistant to degradation and preventing dire consequences on major infrastructures.

Regardless of the research field of interest, molecular adsorption is controlled by the atomistic interactions between the substrate and the adsorbate. The characterization of such interactions is only partially possible by experiments, such as adsorption isotherms~\cite{Kecili-2018} and scanning tunneling microscopy. First-principles calculations play a relevant role in this context as they allow for an accurate description of bond-forming and breaking processes, which is essential to design novel molecular compounds for specific applications.

In recent years, several computational studies on molecular adsorption have been performed with the help of computational tools that allow automatizing the execution of \textit{ab initio} calculations~\cite{DiValentin-2013,Tran-2018-activate,Hong-2009,Montoya-2017,mpinterfaces,Borodin_2015,Jain-2013,Tran-2016}. Most of them are based on a high-throughput approach that allows to calculate the adsorption of simple molecules and molecular fragments in an automatized way and store/retrieve all the generated data in publicly available database~\cite{Tran-2018-activate,Tran-2018-dynamic,Montoya-2017,Boes-2019,Pablo-Garcia-2021}. An example of such a framework is GasPy~\cite{Tran-2018-activate,Tran-2018-dynamic}, based on Fireworks~\cite{fireworks} and Atomate~\cite{atomate} as workflow manager. This approach is highly efficient in computing simultaneously thousands of potential molecules and substrates combinations, resulting very useful for identifying novel potential catalytic materials, a field in which these tools have been widely used~\cite{Tran-2018-activate,Montoya-2017,Boes-2019,Pablo-Garcia-2021}. On the other hand, the platform employed to generate such data is rather complex due to the advanced coding necessary to implement and apply the workflow managers. Non-expert users can encounter difficulties in installing and managing their functionalities. Simpler and lighter programs, such as ASAP~\cite{Wilson-2022} and DockOnSurf~\cite{Marti-2021} have thus been developed. Usually, these programs are based on Python, allowing an easier installation and a more user-friendly handling of the operations~\cite{Wilson-2022,Marti-2021,Steinmann-2022}. The absence of a proper workflow manager can make it more challenging to deal with many configurations and systems. However, specific libraries like the Atomic Simulation Environment (ASE)~\cite{ASE} can help create, generate and handle many systems and configurations simultaneously. 

One of the most critical issue in evaluating molecular adsorption is the sampling and screening of all the possible adsorption configurations that correspond to different orientations of the molecule over different adsorption sites on the substrate. Such configurational space can become huge and almost impossible to explore in its entirety by increasing the complexity of the molecule and the substrate. Therefore, several optimization algorithms have been developed during the years to address this analysis, like metadynamics~\cite{Laio-2002}, Bayesian optimization~\cite{Shahriari-2016,Deshwal-2021}, minima hopping~\cite{Goedecker-2004} and Monte Carlo~\cite{Vignola-2018}. However, linking these programs with complex DFT codes is not trivial~\cite{Todorovic-2019,Rey-2022}. A possible novel technique to predict the most stable molecular adsorption configurations combines machine learning algorithms with DFT calculations, but this work is still in its infancy~\cite{Jung-2022}.

Here we introduce a simple but equally efficient approach to automatically identify the adsorption configurations that have the highest statistical weight, by means of the following steps: i) automatic generation of many adsorption configurations through the molecule rotation and translation with respect to the surface; ii) identification of the most relevant ones by screening their energies through a partial optimization; iii) full structural optimisation of the configurations identified as most relevant. This approach, which is conceptually very simple and with a low level of algorithm sophistication, is anyway very effective as it allows one to easily identify the most relevant adsorption configurations of complex molecules on (reconstructed) surfaces in a relatively short time. Many possible configurations are automatically generated, thus avoiding relying exclusively on user experience and intuition, which can limit the number of configurations analyzed.
The program, \texttt{Xsorb}, here released is written in Python and is based on well-established Python libraries, such as ASE~\cite{ASE} and Pymatgen~\cite{pymatgen}, for the generation of molecular adsorbed systems and on Quantum ESPRESSO (QE)~\cite{QE-orig,QE-adv,QE-exa} for DFT calculations.

The paper is organized as follows: Section~\ref{sec:method} explains how to install \texttt{Xsorb}, how it works and which is its standard workflow during a typical run. Section~\ref{sec:results} shows an examples of \texttt{Xsorb} usage for computing the adsorption of an hydrocarbon molecule, 1-hexene (hexene from now on), over the (110) surface of a bcc crystal, iron in the specific case.

\section{\texttt{Xsorb} structure and execution}
\label{sec:method}

\begin{figure}
\centering
\includegraphics[width=0.5\textwidth]{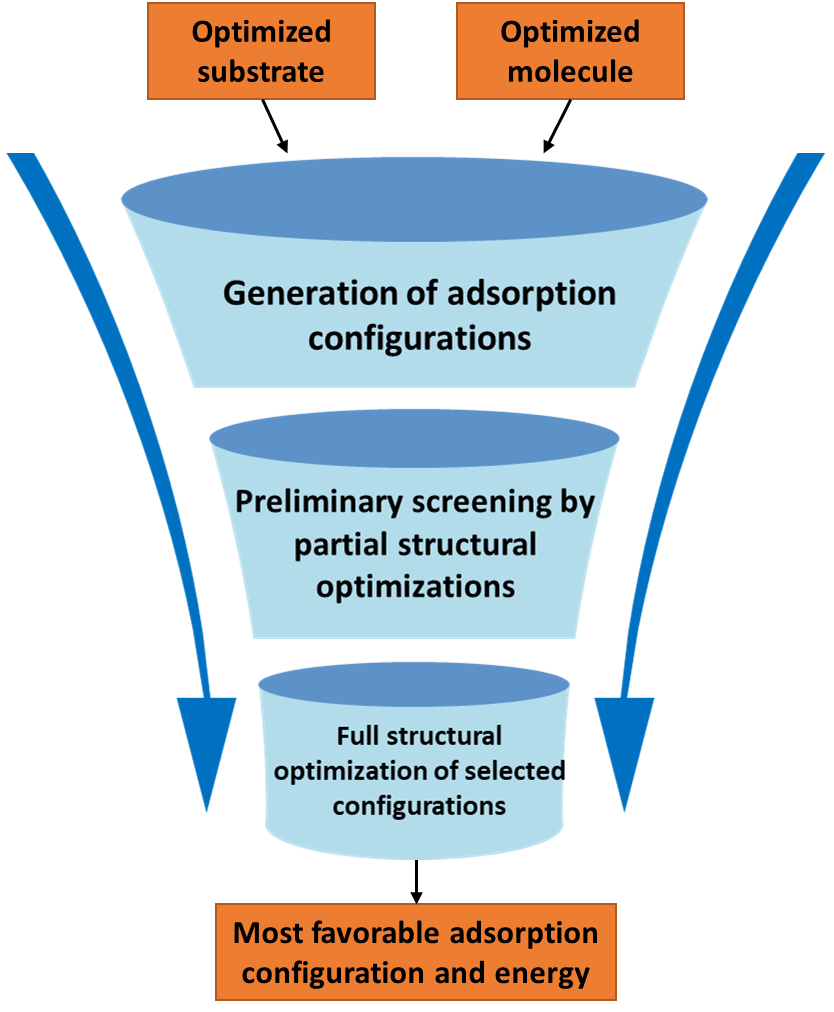}



\caption{Schematic flowchart of \texttt{Xsorb} usage.} \label{fig:flowchart}
\end{figure}

A schematic flowchart of \texttt{Xsorb} functioning is shown in Fig.~\ref{fig:flowchart}.
The main feature of \texttt{Xsorb} is the automatic identification of all the substrate adsorption sites and the generation of all the desired adsorption configurations by applying several molecule rotations for each site. After this initial procedure of input generation, the code automatically launches the calculations. It also provides valuable features to analyze the results, such as extracting the adsorption energies, monitoring the energy evolution during the structural optimization, and generating images or animations of the output files.

\subsection{Defining the setting parameters}
\label{sec:ads_gen}

The main settings for the calculation, such as QE input commands and the list of molecular rotations, must be placed in a \texttt{settings.in} file, located in the folder from which the \texttt{xsorb} command is executed. A template of the \texttt{settings.in} file is provided in the main folder of the repository.

In the following sections, we present a brief outline of the workflow implemented in our code.

A few mandatory flags need to be inserted in the \texttt{settings.in} file to generate the adsorption configurations:

\begin{enumerate}
    \item \texttt{slab\_filename} and \texttt{molecule\_filename} are the names of the two files containing the slab and molecule coordinates, respectively. These optimized structures are retrieved to generate the adsorbed structure.
    \item \texttt{molecule\_axis} is the molecular axis that will be placed along the x direction and considered for rotating the molecule with respect to the substrate. The axis can defined in two ways: i) The user indicates two atoms of the molecule using the variables \texttt{atoms [A1] [A2]}, where \texttt{Ai} are numeric indexes from \texttt{molecule\_filename}, starting from 0. The axis will contain the vector $\mathbf{a} = \mathbf{r}_{A2} - \mathbf{r}_{A1}$. ii) The user explicitly provides the coordinates of the axis vector, \texttt{vector [$v_x$] [$v_y$] [$v_z$]}.
    \item \texttt{selected\_atom\_index} identifies the molecule atom that will be placed in the selected adsorption sites and used as the origin of the reference frame for rotations. The numeric index must be consistent with the order in the \texttt{molecule\_filename}.
    \item \texttt{$x$\_rot\_angles}, \texttt{$y$\_rot\_angles} and \texttt{$z$\_rot\_angles} are the variables that provide a single angle or a list of angles for which the molecule will be rotated around the axis identified by the variable name. The rotations are executed in the following order: first around the $x$ axis, then around the $-y$ axis and finally around the $z$ axis of the reference frame. Following this procedure, all angle combinations are produced. Note that the rotations are always considered along a fixed Cartesian reference frame. This operation implies that, if a rotation \texttt{$y$\_rot\_angles} = 90 degrees is applied, the \texttt{molecule\_axis}, which previously lied along the $x$ axis, now coincides with the $z$ axis. Therefore, to avoid the screening of already sampled molecular orientations around the \texttt{molecule\_axis}, rotations defined by \texttt{$z$\_rot\_angles} are disregarded.
\end{enumerate}

The desired distance between the reference atom of the molecule and the surface adsorption site can be set with the \texttt{screening\_atom\_distance} variable (the default value is 2 \r{A}). To avoid possible overlapping between the molecule and the substrate, an additional variable \texttt{screening\_min\_distance} (with a default value of 1.5 \r{A}) is implemented so all the atoms of the molecules have a minimal vertical displacement from the surface. A final check on the Euclidean distance is done automatically for all the atoms in the molecule to avoid overlapping with the substrate. This control can be essential in atomically rough surfaces such as reconstructed substrates.

\subsection{Generation of adsorption configurations}
\label{sec:ads_site}

\begin{figure*}[htpb]
    \centering
    \begin{subfigure}{0.49\textwidth}
        \centering
        \includegraphics[width=\textwidth]{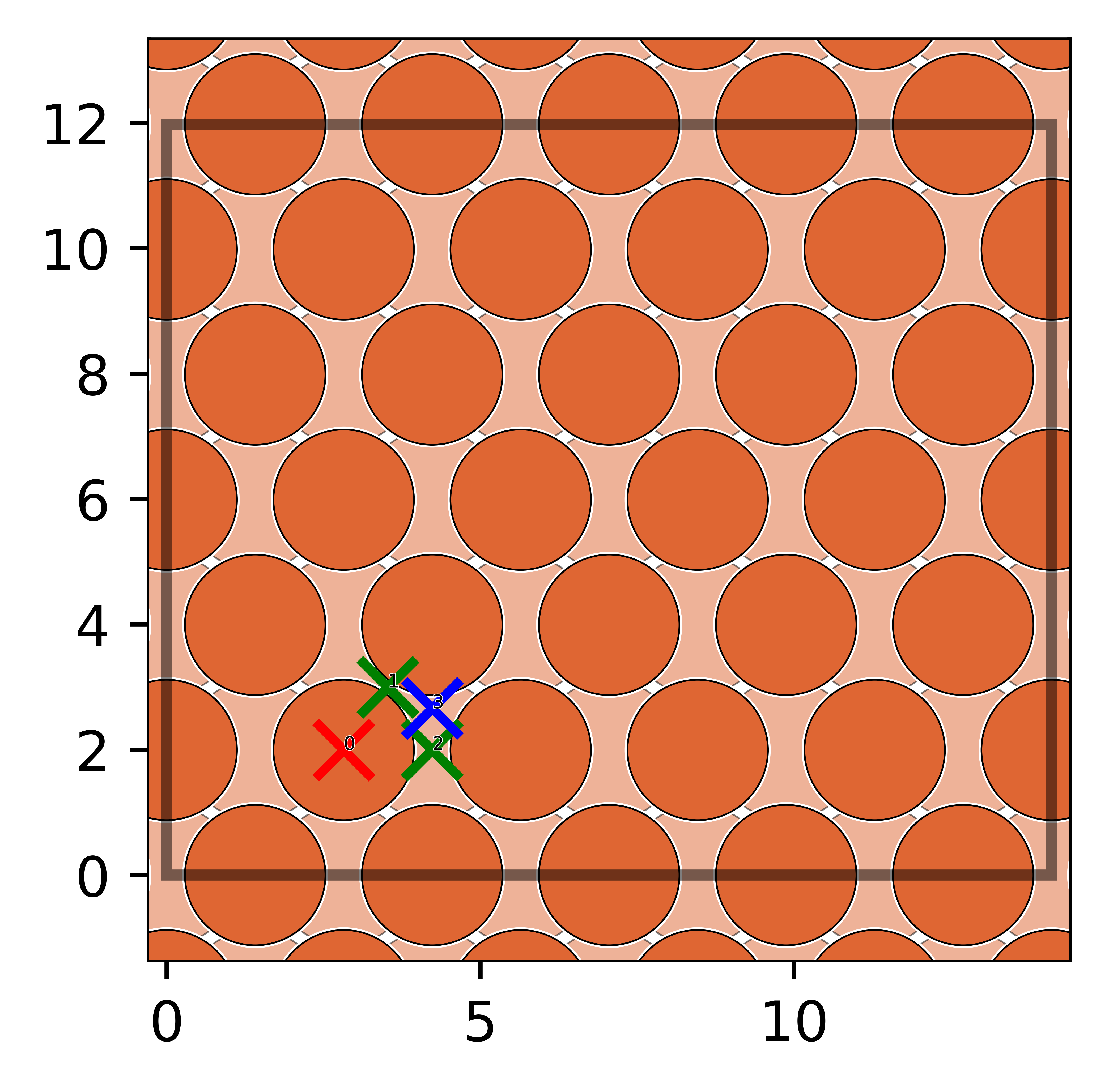}
        \caption{Non-equivalent adsorption sites.}\label{fig:Fe_sites_ineq}
    \end{subfigure}
    \begin{subfigure}{0.49\textwidth}
        \centering
        \includegraphics[width=\textwidth]{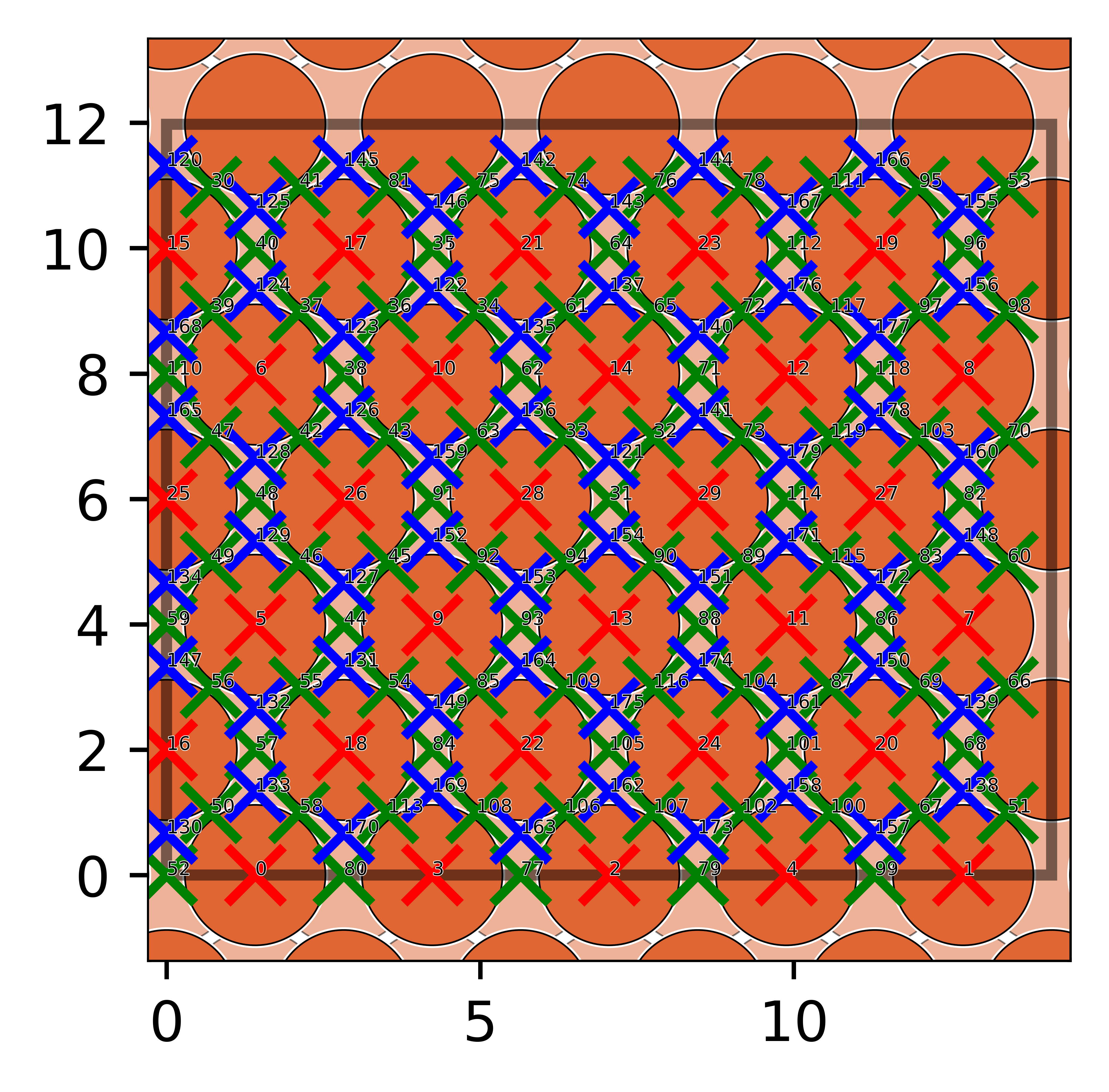}
        \caption{All high-symmetry sites.}\label{fig:Fe_sites_all}
    \end{subfigure}
    \caption{Adsorption sites of the Fe(110) surface. The non-equivalent adsorption sites (Panel a) and all possible sites (Panel b) are shown. The three symmetry types (on-top, bridge, hollow) are colored in red, green and blue, respectively. Fe atoms are colored in different scale of gray to represent the different layers depth within the substrate: the darker (lighter) the color, the shallower (deeper) the layer in which the atom is.}\label{fig:Fe_sites}
\end{figure*}

\texttt{Xsorb} can identify the adsorption sites thanks to the \texttt{AdsorbateSiteFinder} class in the Pymatgen library. This class and its related commands, based on the Delaunay triangulation~\cite{Delaunay-1934} on the topmost surface layer, are very effective in finding all the adsorption sites within the simulation cell. Moreover, for flat and clean surfaces, it also identifies the subset of non-equivalent sites with respect to surface symmetries, as shown in Fig.~\ref{fig:Fe_sites}. However, for more complex substrates, like oxides and doped surfaces, a large number of sites is identified due to reduced surface symmetry. In these cases some manual tuning is necessary to select the most relevant ones, as explained below.

The command \texttt{xsorb -sites} generates a top-view image of the simulation cell where the identified adsorption sites are marked by crosses of different colors to distinguish the different site symmetries (namely, on-top, hollow and bridge). If unexpected results are presented in this image, e.g., some sites are misplaced, it is possible to use the command \texttt{xsorb -sites-all} to obtain all the adsorption sites. In this case, the user should check if there are incorrect or missing on-top sites, adjusting the \texttt{surface\_height} variable in \texttt{settings.in} until the on-top sites are identified correctly. This flag controls the $\Delta z$ value (default 0.9 \r{A}) employed by the \texttt{AdsorbateSiteFinder} class to define the topmost layer of the surface, whose atoms constitute the triangular vertices for the Delaunay triangulation and therefore also determine the positions of all the other adsorption sites.

Another issue that could arise in this automatic identification is the presence of too many (or too few) sites, leading to an over (or under) representation of adsorption configurations. Therefore, the user can increase (reduce) the \texttt{symm\_reduce} variable (with default value of 0.01), used by Pymatgen \texttt{AdsorbateSiteFinder} class to reduce (increase) the number of identified adsorption sites.

\begin{figure*}[htpb]
    \centering
    \begin{subfigure}{0.49\textwidth}
        \centering
        \includegraphics[width=\textwidth]{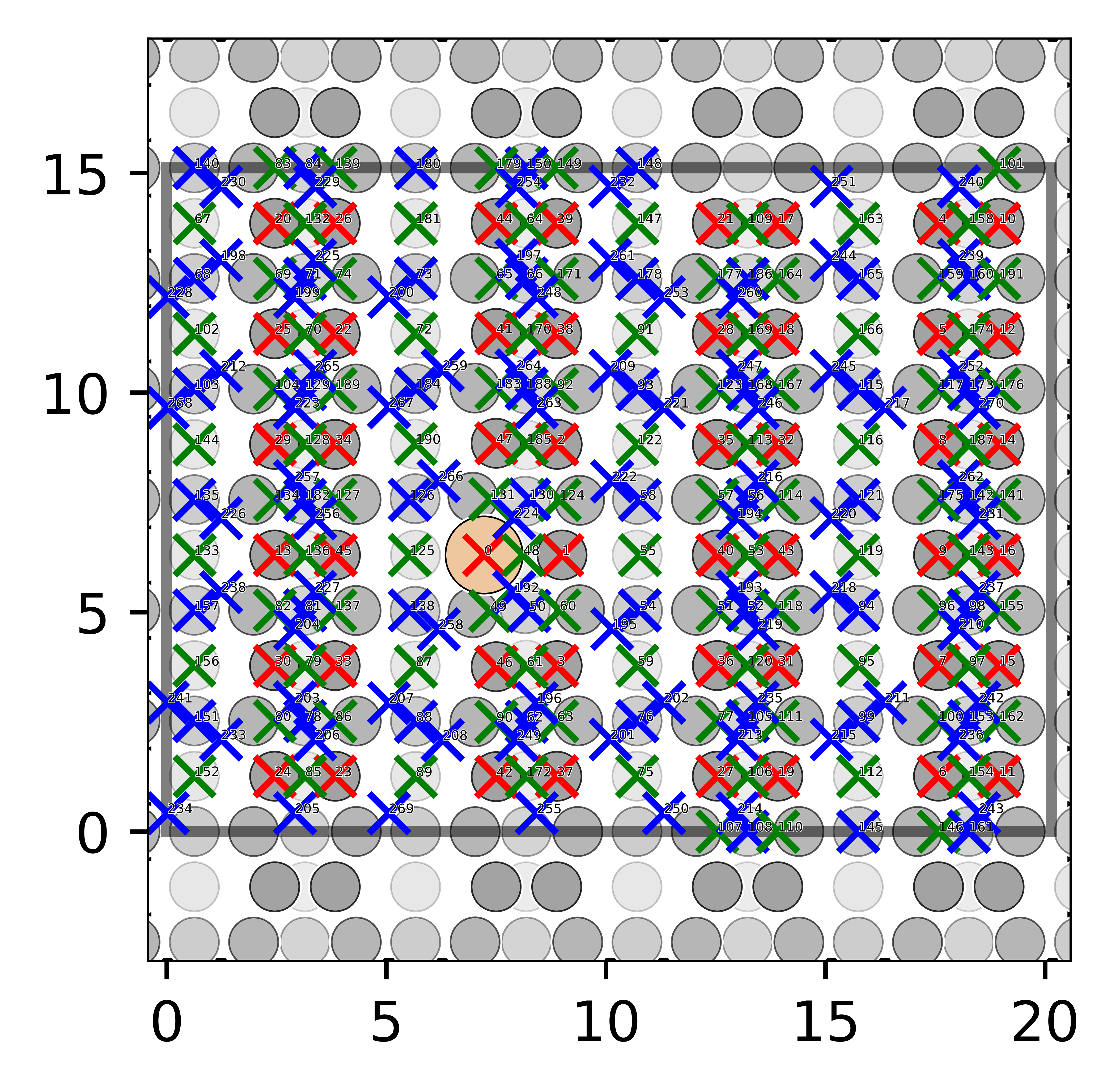}
        \caption{All adsorption sites (\texttt{symm\_reduce} = 0)}\label{fig:C_Si_sites_ineq}
    \end{subfigure}
    \begin{subfigure}{0.49\textwidth}
        \centering
        \includegraphics[width=\textwidth]{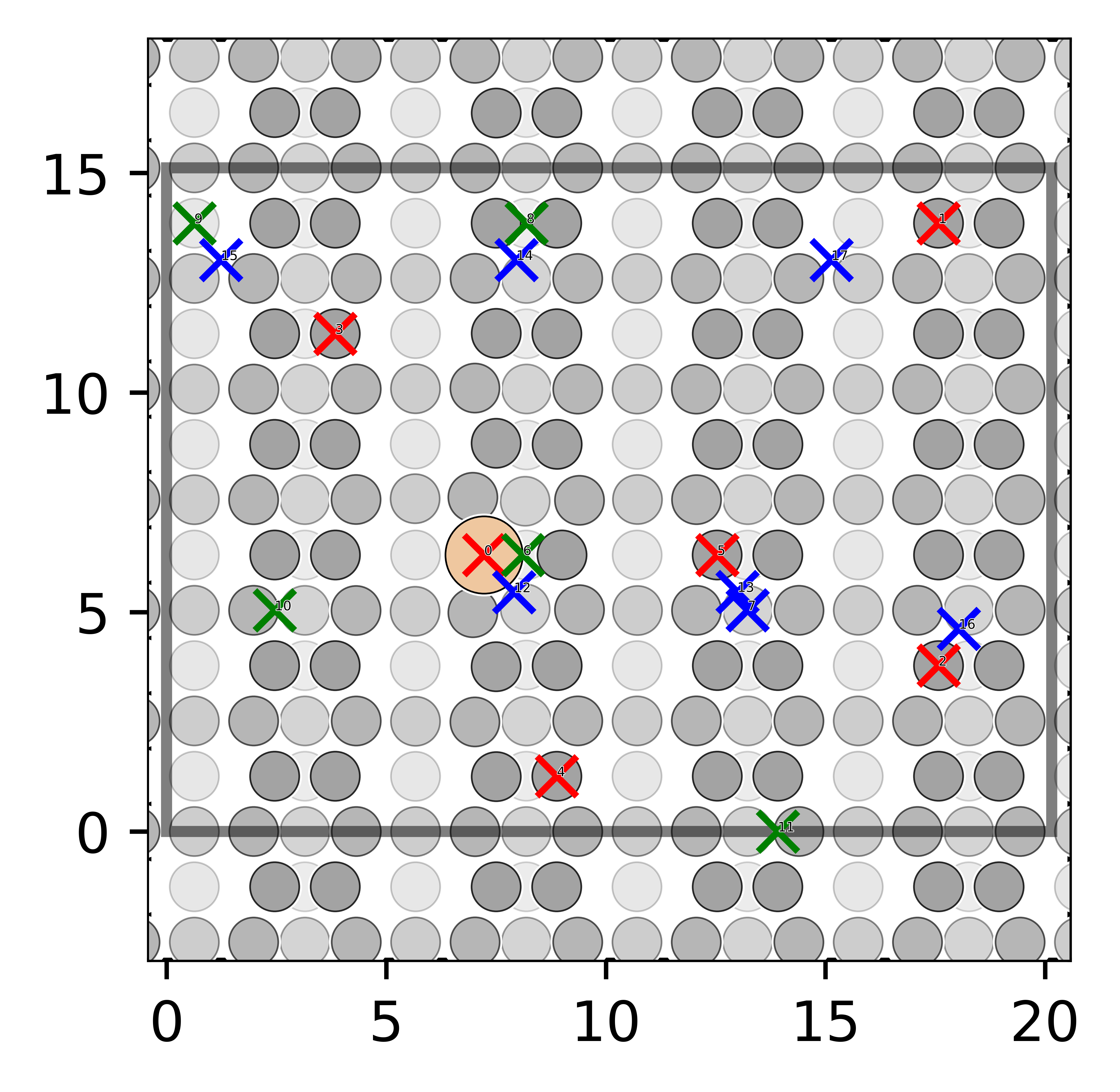}
        \caption{Reduced number of sites (\texttt{symm\_reduce} = 0.25)}\label{fig:C_Si_sites_all}
    \end{subfigure}
    \caption{Adsorption sites of the Si-doped reconstructed C(001) surface. In panel $a$, all possible sites are shown by using \texttt{symm\_reduce} = 0. In panel b, a larger value was used (\texttt{symm\_reduce} = 0.25) and the total number of adsorption sites is significantly reduced. In both cases, each site has a small numeric label, which can be used for the manual site selection. The sites are colored in red, green and blue according to the three symmetry types (on-top, bridge, hollow, respectively). C atoms are colored in different scale of gray to represent the different layers depth within the substrate: the darker (lighter) the color, the shallower (deeper) the layer in which the atom is.}\label{fig:C_Si_sites}
\end{figure*}

Some situations could be even more challenging for site identification. For example, a substrate with a single dopant atom can completely break the symmetry of the surface, producing an excessive number of non-equivalent adsorption sites that could lead to an unfeasible screening of adsorption configurations. Moreover, one might be interested in studying only the sites close to the dopant, neglecting the undoped regions. For these specific cases, the user can impose \texttt{symm\_reduce} = 0 to obtain all adsorption sites and then manually select the required ones by specifying their indexes with the flag \texttt{selected\_sites}. An example of this procedure is represented in Fig.~\ref{fig:C_Si_sites}, where the adsorption sites in a Si-doped diamond surface are represented. Both pictures show the labelling for each adsorption site that can be used for the manual selection.

After the adsorption sites identification, \texttt{Xsorb} generates all the adsorption configurations. Their total number is obtained by multiplying the number of selected adsorption sites by the number of provided rotations along the $x$, $y$ and $z$ axes.

\subsection{Executing DFT calculations}

In order to create the input files to perform the DFT calculations for the generated configurations, it is necessary to specify the computational parameters in the \texttt{settings.in} file using the standard syntax for QE input files (see the \href{https://gitlab.com/triboteam/xsorbed/-/wikis/home}{user guide} for specific details).

As mentioned in the Introduction, two types of calculation are performed: a preliminary screening to identify the most relevant configurations with lowest energies, and a full structural optimization of a subset of these configurations, selected automatically by a pre-defined energy threshold.

The preliminary screening consists in a partial structural optimization, i.e., by fixing higher energy and force thresholds compared to the default values of QE. Typically, we use $5 \times 10^{-3}$ Ry for the energy and $5 \times 10^{-2}$ Ry/bohr for the forces as thresholds. In this way, the user can obtain an estimate of the most promising configurations before launching the full optimizations, at a much lower computational cost.

This screening method, compared to the other techniques approaches described in the Introduction, is completely DFT-based, achieving a greater accuracy in capturing the electronic interactions between the molecule and the substrate. With this technique, we can also immediately identify the presence of dissociation paths over the whole set of tested configurations, since they often occur within the first few optimization steps. 


\subsubsection{Preliminary screening}



The preliminary screening can be started by including the \texttt{jobscript} flag in \texttt{settings.in}. This flag specifies the path of a jobscript compatible with the workload manager (e.g., Slurm) installed on the machine where QE is executed and the command for job submission, e.g. \texttt{sbatch} for Slurm. The calculations can be launched with the following command:

 \begin{lstlisting}[language=bash]
xsorb -s [etot_conv_thr forc_conv_thr]
\end{lstlisting}

where \texttt{etot\_conv\_thr} (\texttt{forc\_conv\_thr}) are optional, with default values of $5\times 10^{-3}$ Ry ($5\times 10^{-2}$ Ry/bohr).

This command submits the jobs and labels each configuration with an integer index starting from 0. It also generates the \texttt{site\_labels.csv} file, schematically represented in Table~\ref{tab:hex_Fe_table}, which contains all the information of each configuration with a label, the applied rotations, the adsorption site coordinates and its symmetry. The latter is expressed in the following way:

\begin{itemize}
    \item \texttt{ontop\_}\textit{\{atomic species\}}, where \textit{atomic species} indicates the species of the surface atom.
    \item \texttt{bridge}\textit{\_\{bridge length\}}, where \textit{bridge length} indicates the distance between the two surface atoms that identify the bridge site.
    \item \texttt{hollow\_c}\textit{\{coordination number\}}, where \textit{coordination\_number} corresponds to the number of first neighboring sites.
\end{itemize}

For each site, a numeric index matching the one in the figure generated with the option \texttt{-sites} is also included.

\subsubsection{Retrieving energies after the preliminary screening} 

Once the preliminary screening is completed, all the energies can be extracted with the command:

\begin{lstlisting}[language=bash]
xsorb -es
\end{lstlisting}

that generates the \texttt{screening\_energies.csv} file containing a label for each adsorption configuration, the symmetry of the adsorption site and the adsorption energies $E_{ads}$ of the screening. If the energy of the isolated system is not available, the file reports the total energies $E_{tot}$ instead.

The adsorption energies $E_{ads}$ can be obtained from the $E_{tot}$ as:

\begin{equation}\label{eq:Eads}
    E_{ads} = E_{tot} - ( E_{slab} + E_{mol} )
\end{equation}

where $E_{slab}$ ($E_{mol}$) is the energy of the system containing the isolated slab (molecule). With this notation, a negative (positive) adsorption energy means a favorable (unfavorable) adsorption configuration. The slab and molecule energies can be either manually specified by the \texttt{E\_slab\_mol} flag contained in the \textit{settings.in} file or provide the QE output files of such configurations to extract this value.

\subsubsection{Full geometrical optimization}

After the screening, it is possible to perform the full geometrical optimization of the most relevant configurations with the following command:

\begin{lstlisting}[language=bash]
xsorb -r [options]
\end{lstlisting}

Without any option, \text{Xsorb} will only perform the optimization for the five configurations with lowest energy obtained from the preliminary screening. This number can be changed by employing the option \texttt{--n NUMBER}. Alternatively, the configurations that will be optimized can be selected by specifying an energy threshold with the option \texttt{--t ENERGY}. In this way, all the configurations with an \texttt{ENERGY} (in eV) above the identified minimum will be considered for the full optimization. It is also possible to exclude configurations that would be otherwise be included with the two aforementioned selection methods. The user can do that by using the flag \texttt{--exclude}, as in the following example:

\begin{lstlisting}[language=bash]
xsorb -r --t 0.5 --exclude 1 3 7
\end{lstlisting}

where \texttt{Xsorb} launches the optimization of all configurations within 0.5 eV from the energy minimum of the screening, excluding the ones labelled with 1, 3 and 7. 
Alternatively, the user can provide a list of specific configurations to be optimized using the flag \texttt{--i}, specifying their labels.

\subsubsection{Retrieving final energies after optimizations}

After performing the geometrical optimizations, the user can retrieve their final energies with the command:

\begin{lstlisting}[language=bash]
xsorb -er
\end{lstlisting}

That creates the \texttt{relax\_energies.csv} file, which contains a label for each adsorption configuration, the symmetry of the adsorption site, the total energies $E_{tot}$ of the screening and the total energy of the optimized configuration. In the same way as for the screening, \texttt{Xsorb} can print the optimized adsorption energies when the user provides the energies of the isolated slab and molecule.

\section{Installation and execution of \texttt{Xsorb}}

The installation procedure for \texttt{Xsorb} is easy and straightforward since it is Python-based. It is first necessary to download the program by cloning this repository into the desired local machine:  

\begin{lstlisting}[language=bash]
git clone
https://gitlab.com/triboteam/xsorbed.git
\end{lstlisting}

Once the download is completed, it is necessary to go to the \texttt{xsorbed} main folder and run: 

\begin{lstlisting}[language=bash]
bash install.sh
\end{lstlisting}

to add the executable in the \texttt{PATH} variable. With this operation, the user can launch the program from any folder. The next step is to install the required dependencies stored in the \texttt{requirements.txt} file. This step can be performed with the command:

\begin{lstlisting}[language=bash]
pip install -r requirements.txt
\end{lstlisting}

Further details for the installation procedure, like adding the POV-Ray visualization tool, can be found at the \href{https://gitlab.com/triboteam/xsorbed/-/wikis/home}{user guide}.
\texttt{Xsorb} is interfaced with QE to perform the DFT calculations. It is, therefore, necessary to compile and install the \texttt{pw.x} executable before launching the calculations.

After the installation, the program works through a command-line interface (CLI) by running:  

\begin{lstlisting}[language=bash]
xsorb [command] [parameters]
\end{lstlisting}

Where \texttt{command} specifies which operation to perform and \texttt{parameters} are additional parameters available for specific commands.  

\section{Example of \texttt{Xsorb} use}
\label{sec:results}

\begin{figure}
    \centering
    \includegraphics[width=\columnwidth]{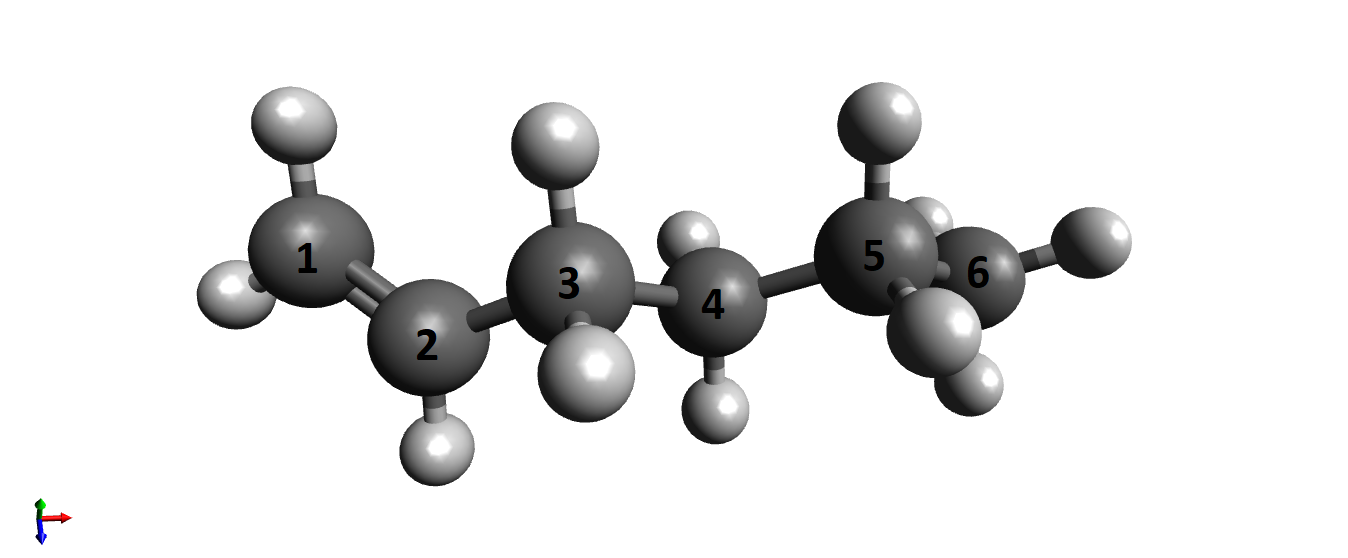}
    \caption{1-hexene molecule, presenting double bond between the carbon atoms labelled 1 and 2.}
    \label{fig:hex_molecule}
\end{figure}

After explaining in detail how \texttt{Xsorb} works, we will present in the following sections the results obtained for the adsorption of a hydrocarbon, hexene, represented in Fig.~\ref{fig:hex_molecule}, over a simple transition metal surface, namely Fe(110).

The structure for the hexene was initially taken from the PubChem database~\cite{hexene-PubChem}, while the two surfaces were generated using Pymatgen. We performed a geometrical optimization separately for each system.

For the DFT calculations, we used an energy cutoff of 40 Ry and a k-point sampling in $\Gamma$. The choice of using only the $\Gamma$-point was justified by the large size of the cells ($14.1 \times 12.0 \times 24.0$ \r{A}$^3$ for Fe, containing 120 atoms, and $20.2 \times 15.1 \times 24.0$ \r{A}$^3$ for C, containing 480 C atoms), since denser k-point grids produced total energy differences lower than 1 meV/atom. van der Waals corrections were applied using the Grimme-D2~\cite{Grimme-2006} scheme implemented in Quantum Espresso.

\subsection{Hexene adsorption on Fe(110)}

\subsubsection{Generation of adsorption configurations}

In the case of hexene, the molecule atoms that are more likely to interact with the surface are the two carbon atoms sharing the double bond. Thus, we selected the C atom with only one hydrogen, labelled 2 in Fig.~\ref{fig:hex_molecule}, as the reference atom.

Even if hexene has a C$_1$ point group symmetry, i.e., the only symmetrical operation available is the identity, it is a quite elongated molecule. Thus, to define the \texttt{molecule\_axis}, two atoms along this preferential direction were taken. In particular, we chose the C atoms labelled as 2 and 6 in Fig.~\ref{fig:hex_molecule} to identify this axis.

Finally, to generate all the adsorption structures, it is necessary to define the molecular rotations. This procedure partially relies on user intuition, but a sensible choice can be done by examining both surface symmetries and the molecular structure. 
Since hexene has no internal rotational symmetries, we started by considering the relevant rotations along the $x$ axis, namely 0, 90 and 180 degrees. Moreover, due to the molecular elongation, we selected two additional rotations around the $y$ axis: 0 degrees, which corresponds to the molecule parallel to the surface, and 90 degrees, corresponding to a vertical orientation.
We considered the surface symmetries for the rotations around the $z$ axis and chose three orientations equal to 0, 54.7 and 90 degrees. For what concerns the vertical distance between the surface and the molecule, we decided to keep the \texttt{screening\_atom\_distance} and \texttt{screeening\_min\_distance} variables to their default values, i.e., 2 and 1.5 \r{A}, respectively. We performed the automatic identification of the adsorption sites through the \texttt{xsorb -sites} command. As shown in Fig.~\ref{fig:Fe_sites}, the Fe(110) is a flat and symmetric surface, and the default settings were sufficient to identify all the non-equivalent adsorption sites correctly.

A visual representation of the adsorption configurations generated for the on-top site can be seen in Fig.~\ref{fig:Fe_screening_top}. The complete set of adsorption configurations can be found in the Supplementary Materials.

\begin{figure*}[htpb]
    \centering
    \begin{subfigure}{0.32\textwidth}
        \centering
        \includegraphics[width=\textwidth]{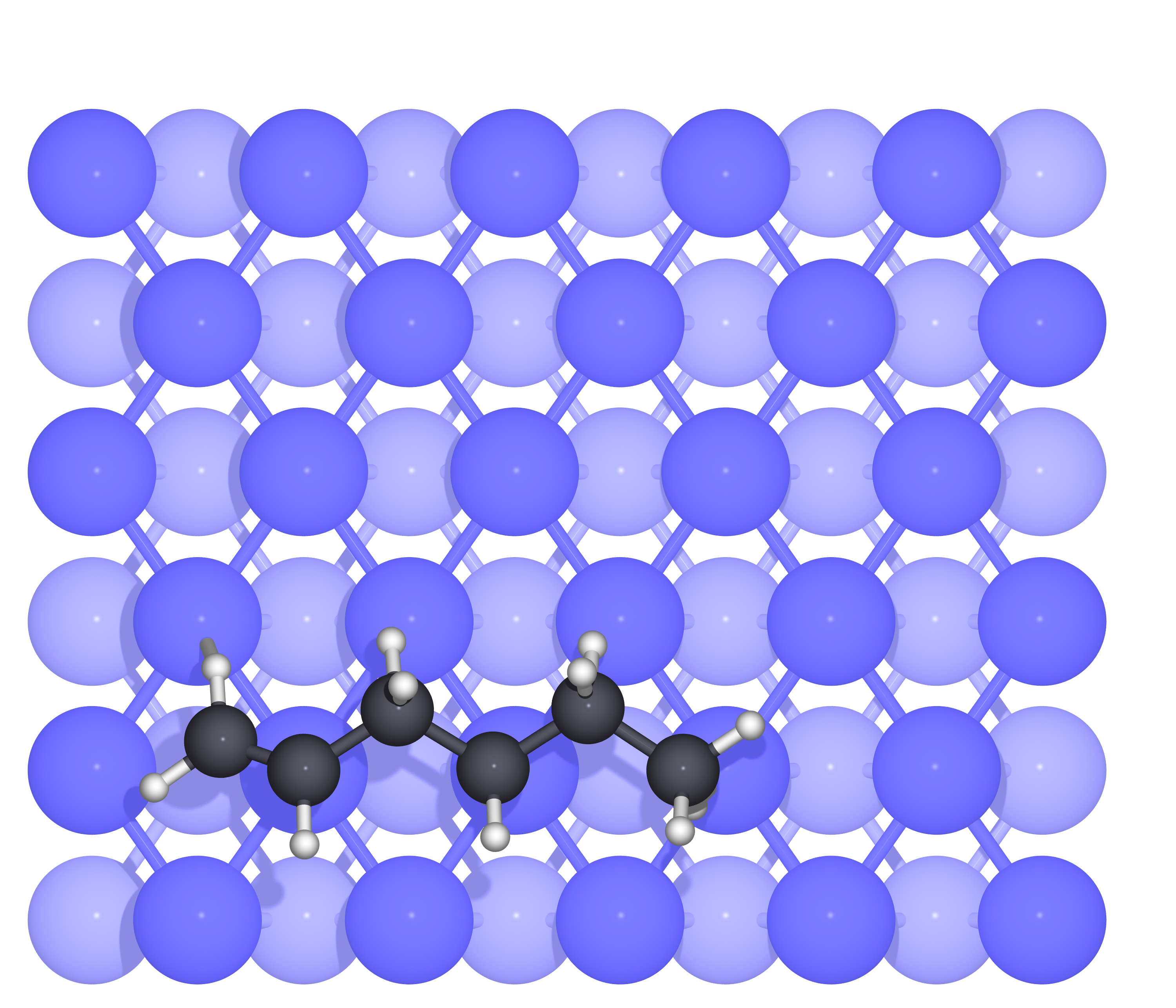}
        \caption{Label: 0}\label{fig:Fe_screening0_top}
    \end{subfigure}
    \hfill
    \begin{subfigure}{0.32\textwidth}
        \centering
        \includegraphics[width=\textwidth]{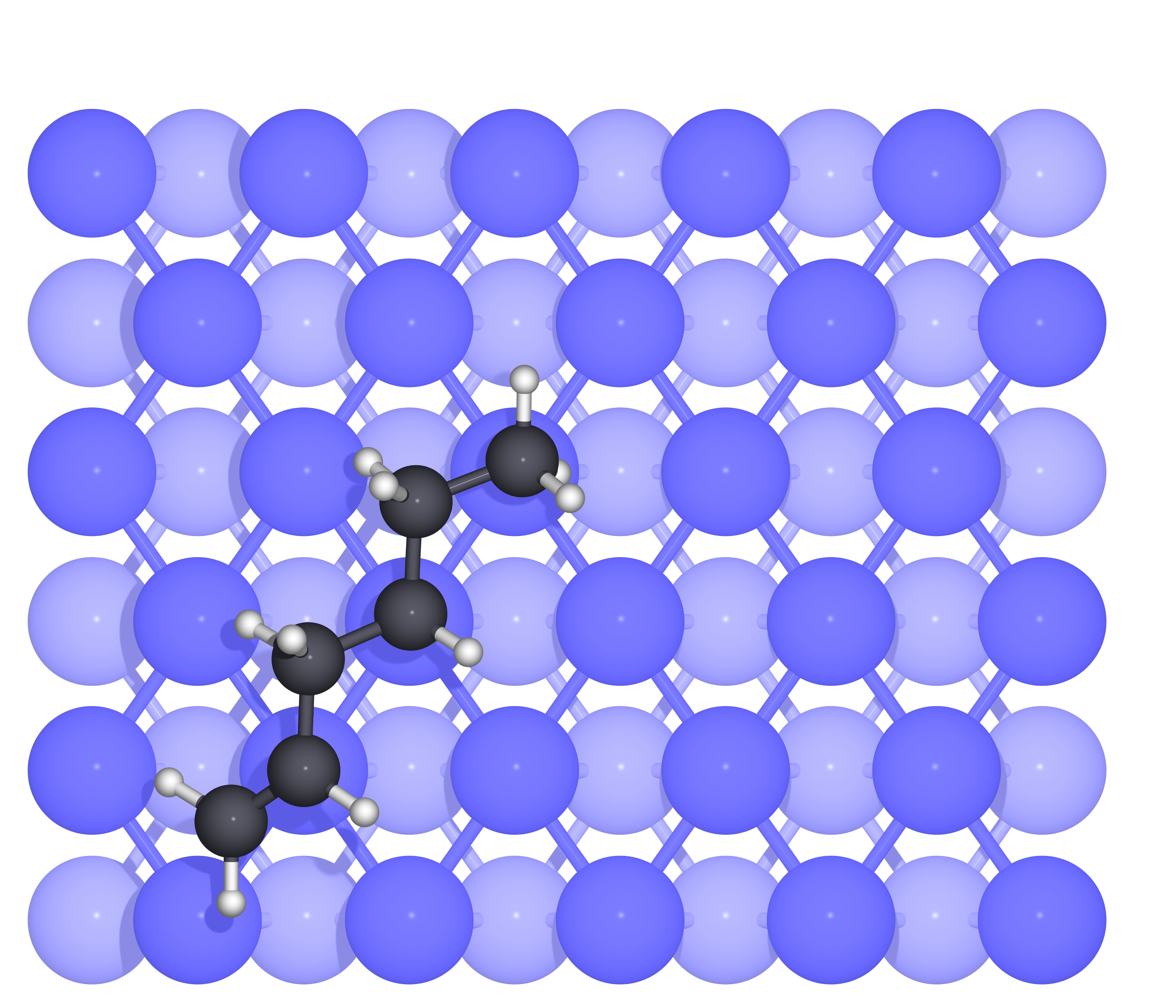}
        \caption{Label: 4}\label{fig:Fe_screening4_top}
    \end{subfigure}
    \hfill
    \begin{subfigure}{0.32\textwidth}
        \centering
        \includegraphics[width=\textwidth]{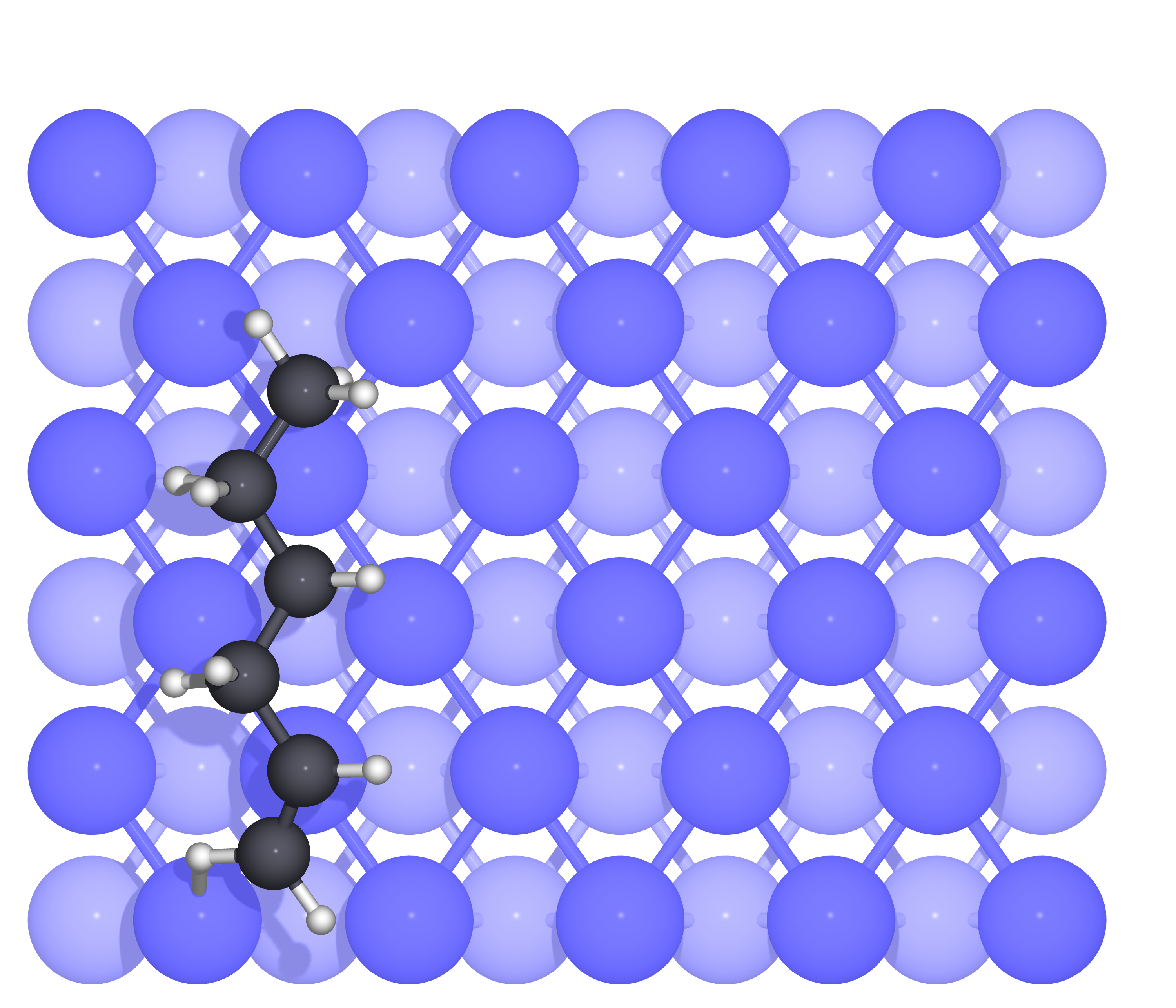}
        \caption{Label: 8}\label{fig:Fe_screening8_top}
    \end{subfigure}
    \hfill
    \begin{subfigure}{0.32\textwidth}
        \centering
        \includegraphics[width=\textwidth]{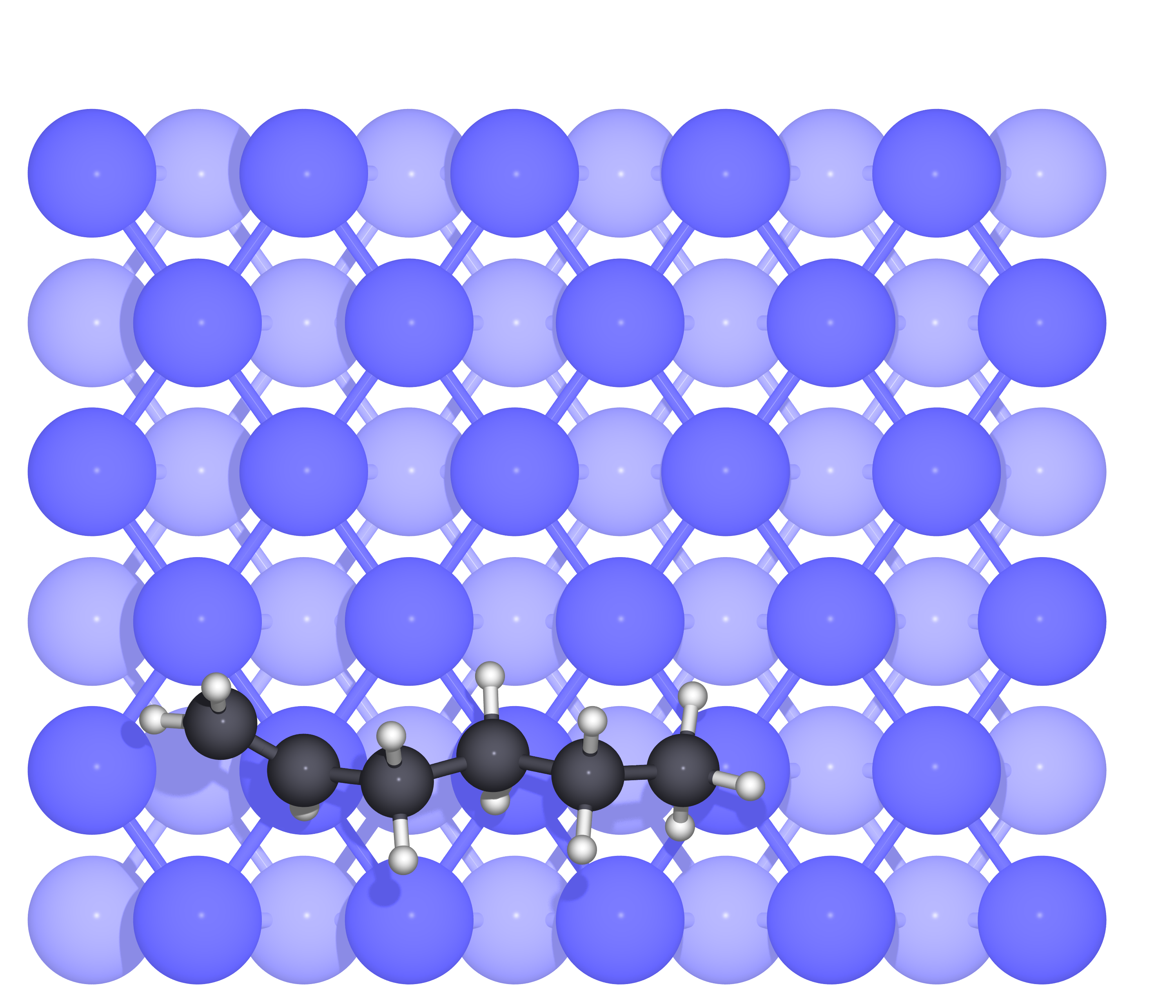}
        \caption{Label: 12}\label{fig:Fe_screening12_top}
    \end{subfigure}
    \begin{subfigure}{0.32\textwidth}
        \centering
        \includegraphics[width=\textwidth]{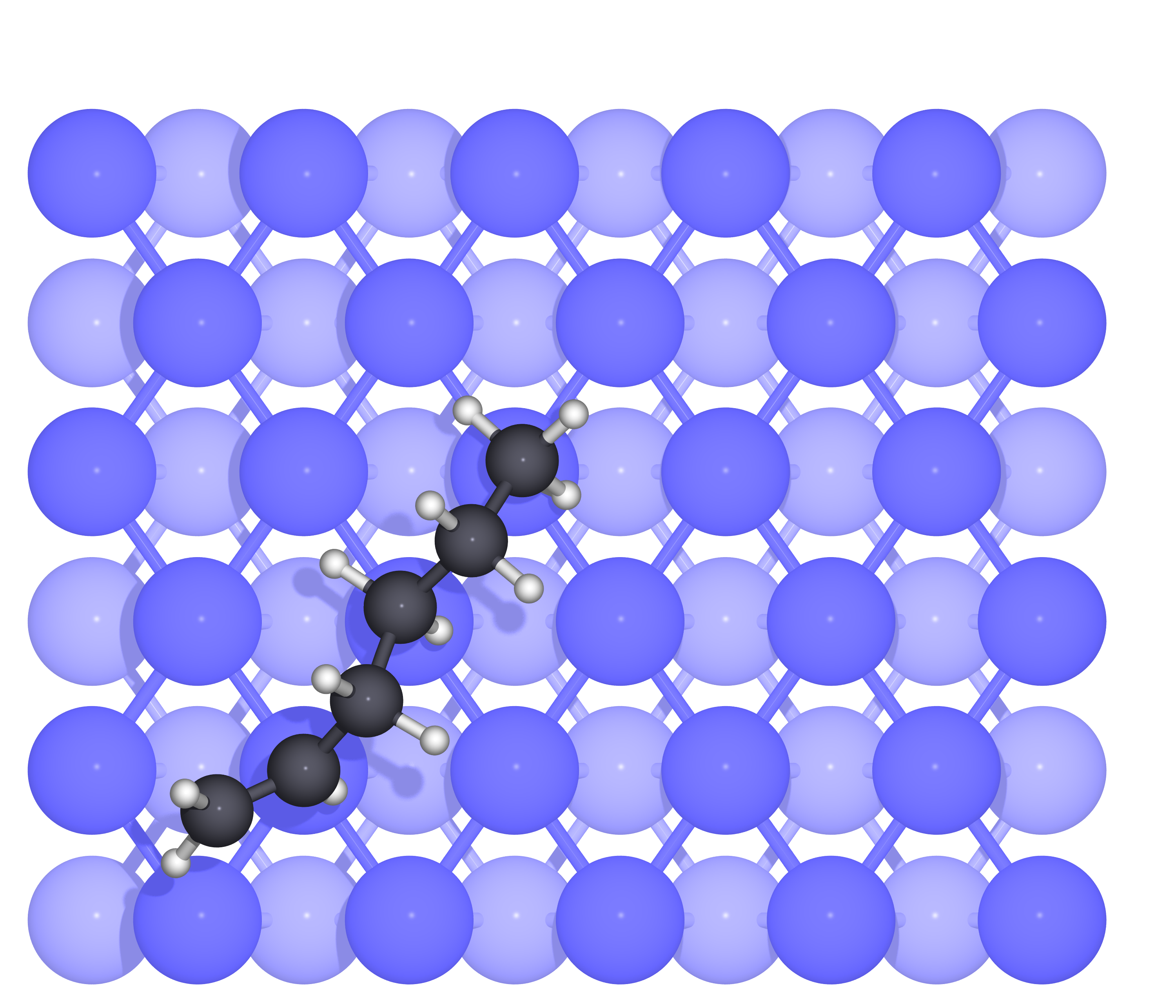}
        \caption{Label: 16}\label{fig:Fe_screening16_top}
    \end{subfigure}
    \begin{subfigure}{0.32\textwidth}
        \centering
        \includegraphics[width=\textwidth]{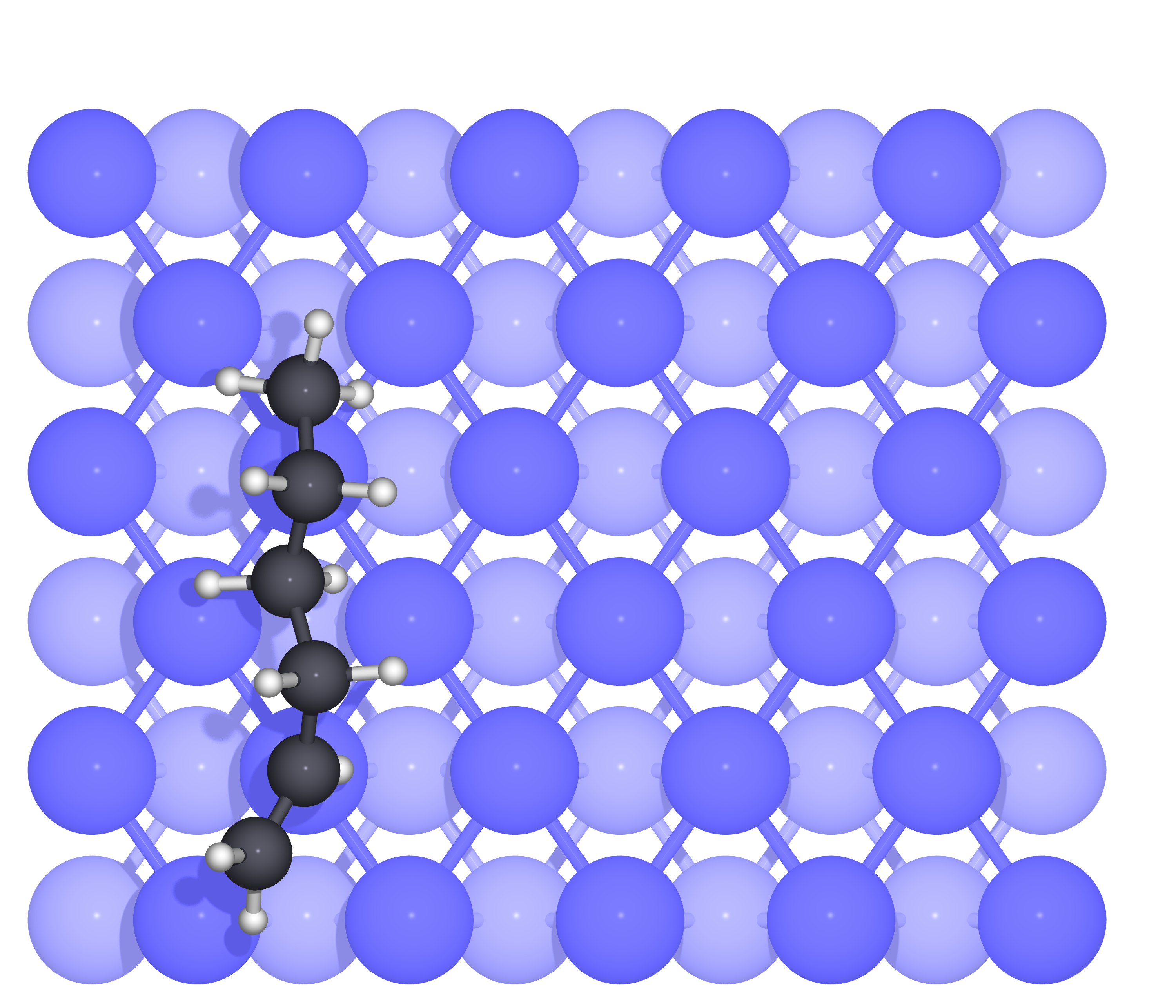}
        \caption{Label: 20}\label{fig:Fe_screening20_top}
    \end{subfigure}
    \begin{subfigure}{0.32\textwidth}
        \centering
        \includegraphics[width=\textwidth]{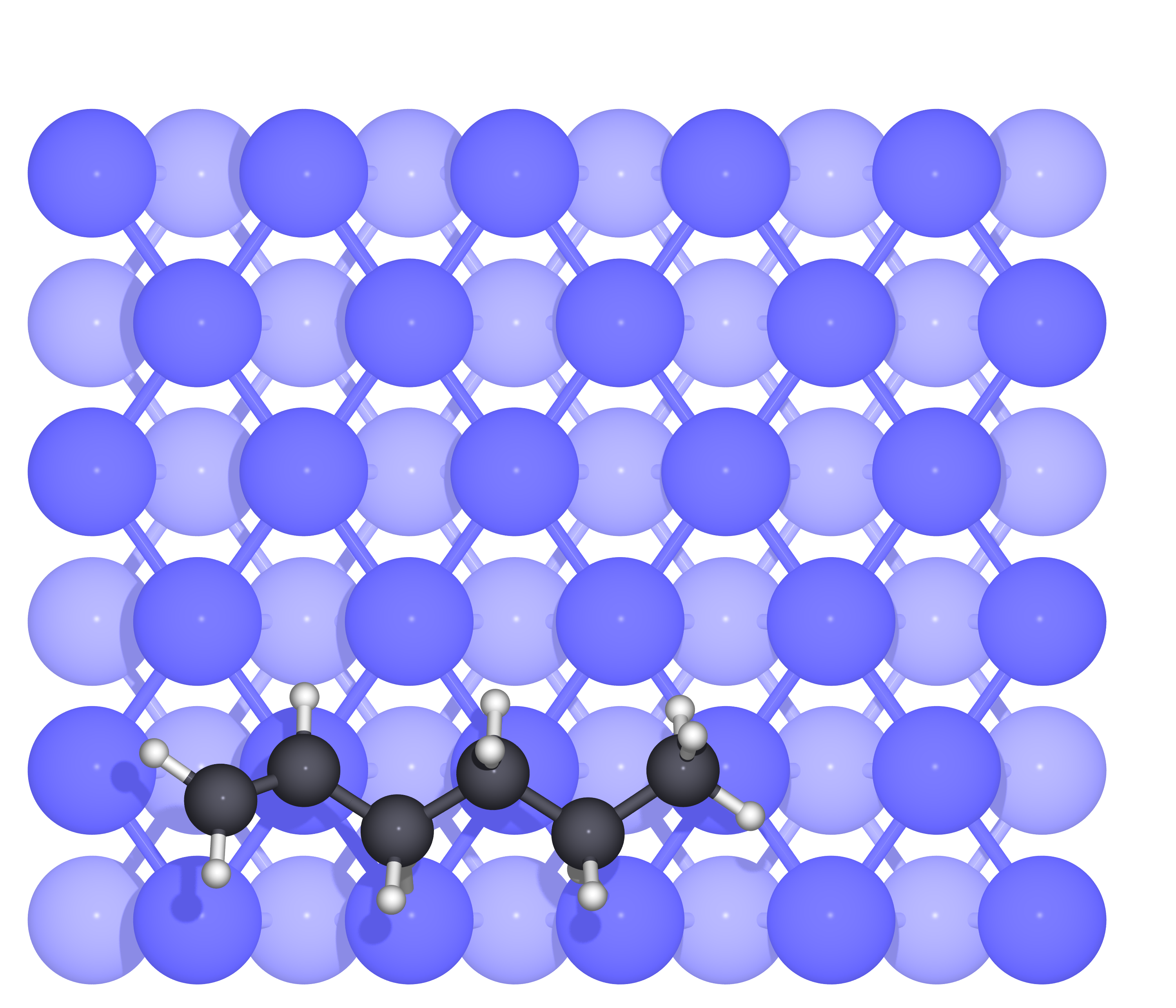}
        \caption{Label: 24}\label{fig:Fe_screening24_top}
    \end{subfigure}
    \begin{subfigure}{0.32\textwidth}
        \centering
        \includegraphics[width=\textwidth]{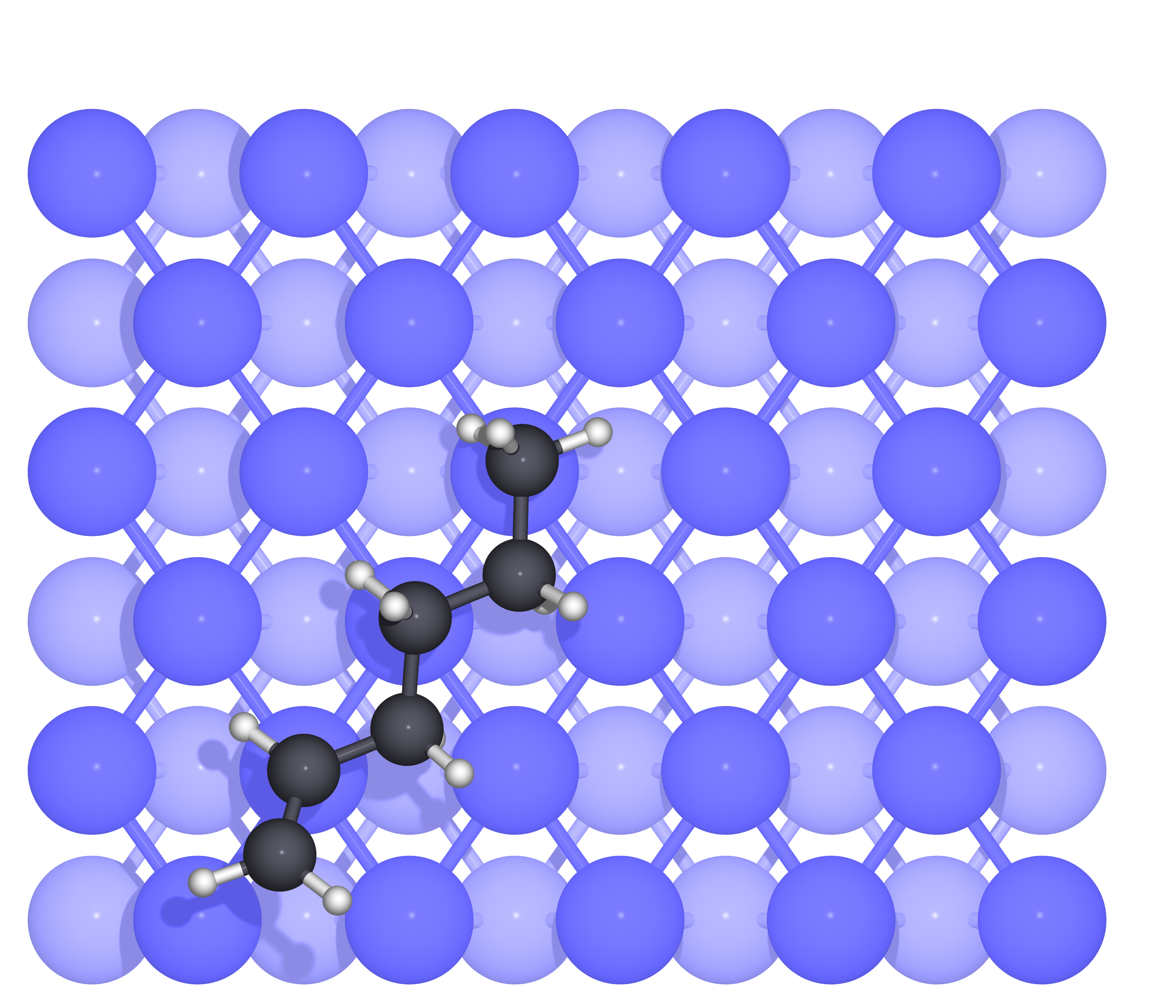}
        \caption{Label: 28}\label{fig:Fe_screening28_top}
    \end{subfigure}
    \begin{subfigure}{0.32\textwidth}
        \centering
        \includegraphics[width=\textwidth]{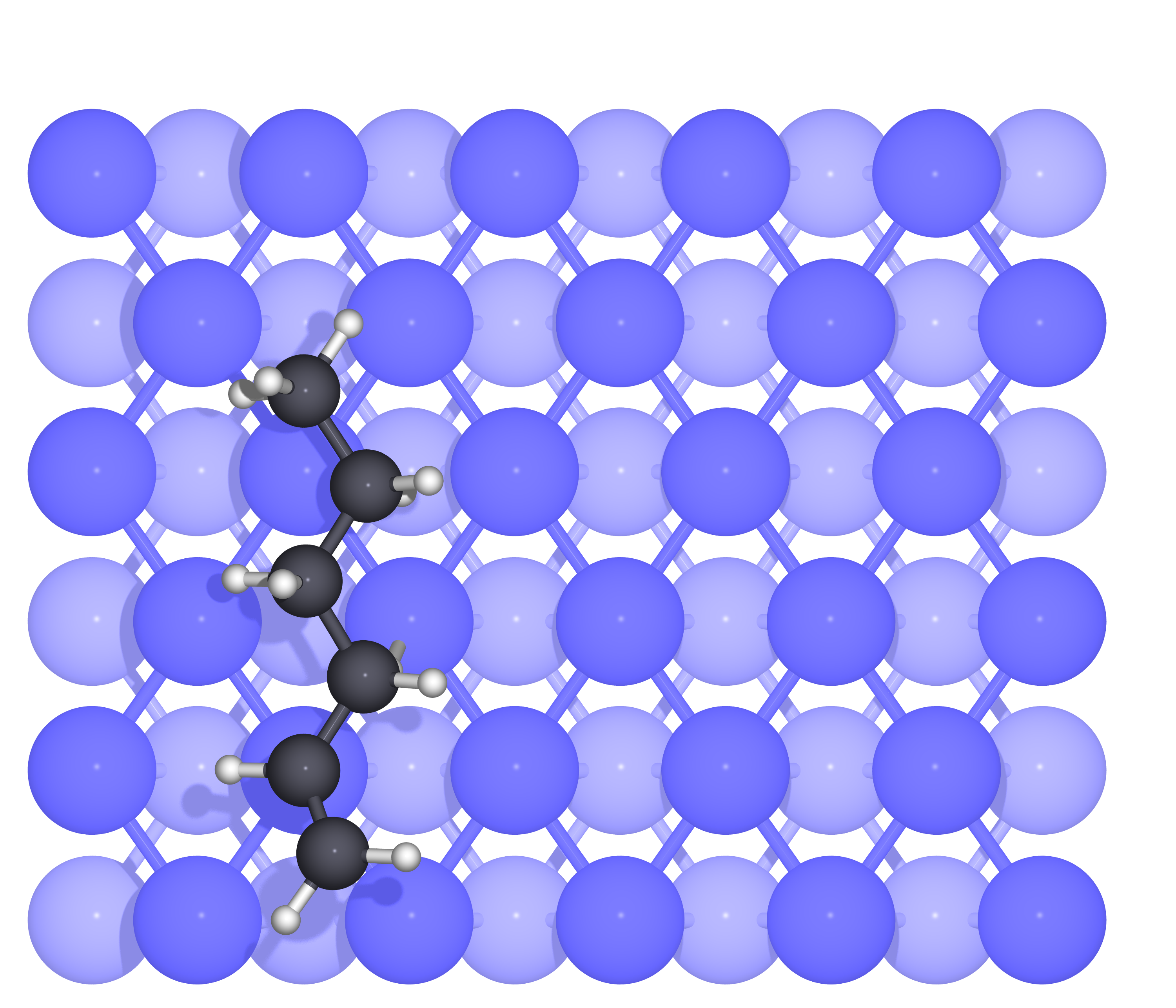}
        \caption{Label: 32}\label{fig:Fe_screening32_top}
    \end{subfigure}
    \begin{subfigure}{0.32\textwidth}
        \centering
        \includegraphics[width=\textwidth]{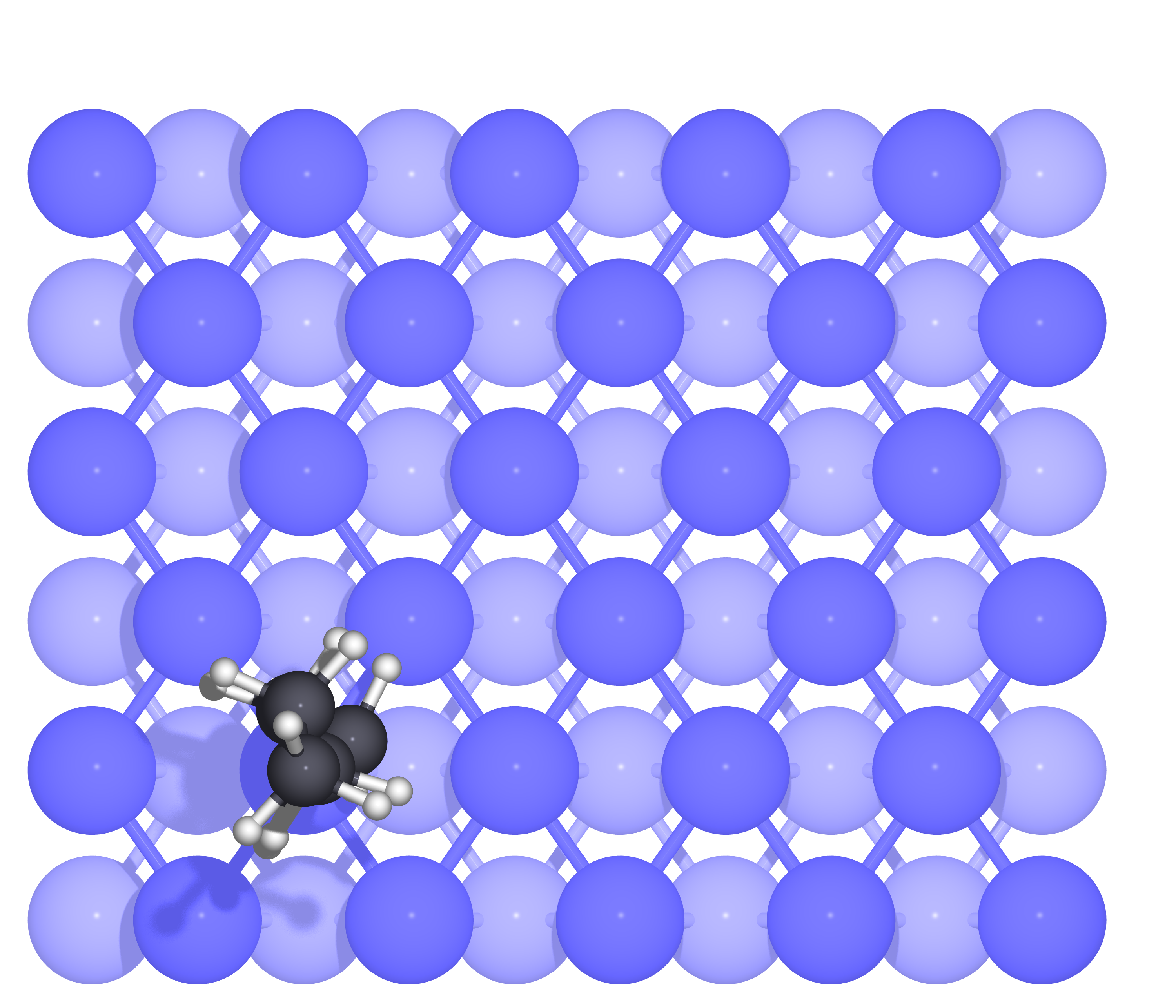}
        \caption{Label: 36}\label{fig:Fe_screening36_top}
    \end{subfigure}
    \caption{Adsorption configuration for hexene on Fe(110) for the preliminary screening. The reference atom is positioned at the on-top site, and all the selected molecular rotations are represented. Other 30 configurations are explored by positioning the reference atom in bridge and hollow sites. The configurations are labelled according to the labels in Tab.~\ref{tab:hex_Fe_table}}
    \label{fig:Fe_screening_top}
\end{figure*}

\subsubsection{Preliminary screening and full geometrical optimization}

After the generation of all the adsorption structures, we performed the preliminary screening. The most relevant data obtained for each configuration are reported in Tab.~\ref{tab:hex_Fe_table}. 
Regarding the vertical distance between the reference atom and the surface, $d$, it is possible to notice that the program automatically moved the molecule upwards to satisfy the minimum distance requirement and avoid any overlap. Moreover, since this operation increased the vertical distance, it avoided the dissociation of hydrogen atoms which are too close to the substrate.

\begin{table}[htbp]
    \centering
    \caption{Data for the preliminary screening of hexene on Fe(110). The data reported in each column corresponds to: the numerical label of the configuration, rotations in degrees along $x$ ($x_{rot}$), $y$ ($y_{rot}$) and $z$ ($z_{rot}$) axis, the index (as in Fig.\ref{fig:Fe_sites}) and the symmetry of the adsorption site, the distance ($d$) between the reference atom and the surface (in \r{A}) and the adsorption energies ($E_{ads}$) of the preliminary screening and of the full optimization, in eV. The systems are sorted according to their screening adsorption energy.}
    \label{tab:hex_Fe_table}
\resizebox{\columnwidth}{!}{%
\begin{tabular}{c c c c c c c c c c}
{Label} & $x_{rot}$ & $y_{rot}$ & $z_{rot}$ &  Symm. site & $d$ &  Screen. $E_{ads}$ & Opt. $E_{ads}$\\
\hline
0 & 0 & 0 & 0 & 0 ontop\_Fe & 2.76 & -1.45 & -1.83 \\
4 & 0 & 0 & 54.7 & 0 ontop\_Fe & 2.76 & -1.38 & -1.81 \\
8 & 0 & 0 & 90 & 0 ontop\_Fe & 2.76 & -1.13 & -1.78 \\
1 & 0 & 0 & 0 & 1 bridge\_2.44 & 2.76 & -0.36 & -1.80 \\
11 & 0 & 0 & 90 & 3 hollow\_c3 & 2.76 & -0.34 & -1.68 \\
6 & 0 & 0 & 54.7 & 2 bridge\_2.82 & 2.76 & -0.25 & - \\
10 & 0 & 0 & 90 & 2 bridge\_2.82 & 2.76 & -0.23 & - \\
5 & 0 & 0 & 54.7 & 1 bridge\_2.44 & 2.76 & -0.20 & - \\
9 & 0 & 0 & 90 & 1 bridge\_2.44 & 2.76 & -0.19 & - \\
27 & 180 & 0 & 0 & 3 hollow\_c3 & 2.70 & -0.19 & - \\
7 & 0 & 0 & 54.7 & 3 hollow\_c3 & 2.76 & -0.18 & - \\
3 & 0 & 0 & 0 & 3 hollow\_c3 & 2.76 & -0.16 & - \\
30 & 180 & 0 & 54.7 & 2 bridge\_2.82 & 2.70 & -0.14 & - \\
34 & 180 & 0 & 90 & 2 bridge\_2.82 & 2.70 & -0.13 & - \\
33 & 180 & 0 & 90 & 1 bridge\_2.44 & 2.70 & -0.09 & - \\
16 & 90 & 0 & 54.7 & 0 ontop\_Fe & 2.49 & -0.08 & - \\
12 & 90 & 0 & 0 & 0 ontop\_Fe & 2.49 & -0.07 & - \\
25 & 180 & 0 & 0 & 1 bridge\_2.44 & 2.70 & -0.07 & - \\
2 & 0 & 0 & 0 & 2 bridge\_2.82 & 2.76 & -0.06 & - \\
35 & 180 & 0 & 90 & 3 hollow\_c3 & 2.70 & -0.05 & - \\
31 & 180 & 0 & 54.7 & 3 hollow\_c3 & 2.70 & -0.03 & - \\
24 & 180 & 0 & 0 & 0 ontop\_Fe & 2.70 & -0.03 & - \\
29 & 180 & 0 & 54.7 & 1 bridge\_2.44 & 2.70 & -0.03 & - \\
32 & 180 & 0 & 90 & 0 ontop\_Fe & 2.70 & -0.02 & - \\
26 & 180 & 0 & 0 & 2 bridge\_2.82 & 2.70 & -0.02 & - \\
38 & 0 & 90 & 0 & 2 bridge\_2.82 & 3.50 & 0.01 & - \\
28 & 180 & 0 & 54.7 & 0 ontop\_Fe & 2.70 & 0.03 & - \\
37 & 0 & 90 & 0 & 1 bridge\_2.44 & 3.50 & 0.04 & - \\
21 & 90 & 0 & 90 & 1 bridge\_2.44 & 2.49 & 0.07 & - \\
36 & 0 & 90 & 0 & 0 ontop\_Fe & 3.50 & 0.12 & - \\
23 & 90 & 0 & 90 & 3 hollow\_c3 & 2.49 & 0.13 & - \\
20 & 90 & 0 & 90 & 0 ontop\_Fe & 2.49 & 0.14 & - \\
18 & 90 & 0 & 54.7 & 2 bridge\_2.82 & 2.49 & 0.14 & - \\
39 & 0 & 90 & 0 & 3 hollow\_c3 & 3.50 & 0.16 & - \\
22 & 90 & 0 & 90 & 2 bridge\_2.82 & 2.49 & 0.17 & - \\
13 & 90 & 0 & 0 & 1 bridge\_2.44 & 2.49 & 0.18 & - \\
15 & 90 & 0 & 0 & 3 hollow\_c3 & 2.49 & 0.23 & - \\
19 & 90 & 0 & 54.7 & 3 hollow\_c3 & 2.49 & 0.31 & - \\
14 & 90 & 0 & 0 & 2 bridge\_2.82 & 2.49 & 0.35 & - \\
17 & 90 & 0 & 54.7 & 1 bridge\_2.44 & 2.49 & 0.40 & - \\
\hline
\end{tabular}
}
\end{table}

\begin{figure}[htpb]
    \centering
    \begin{subfigure}{0.23\textwidth}
        \centering
        \includegraphics[width=\textwidth]{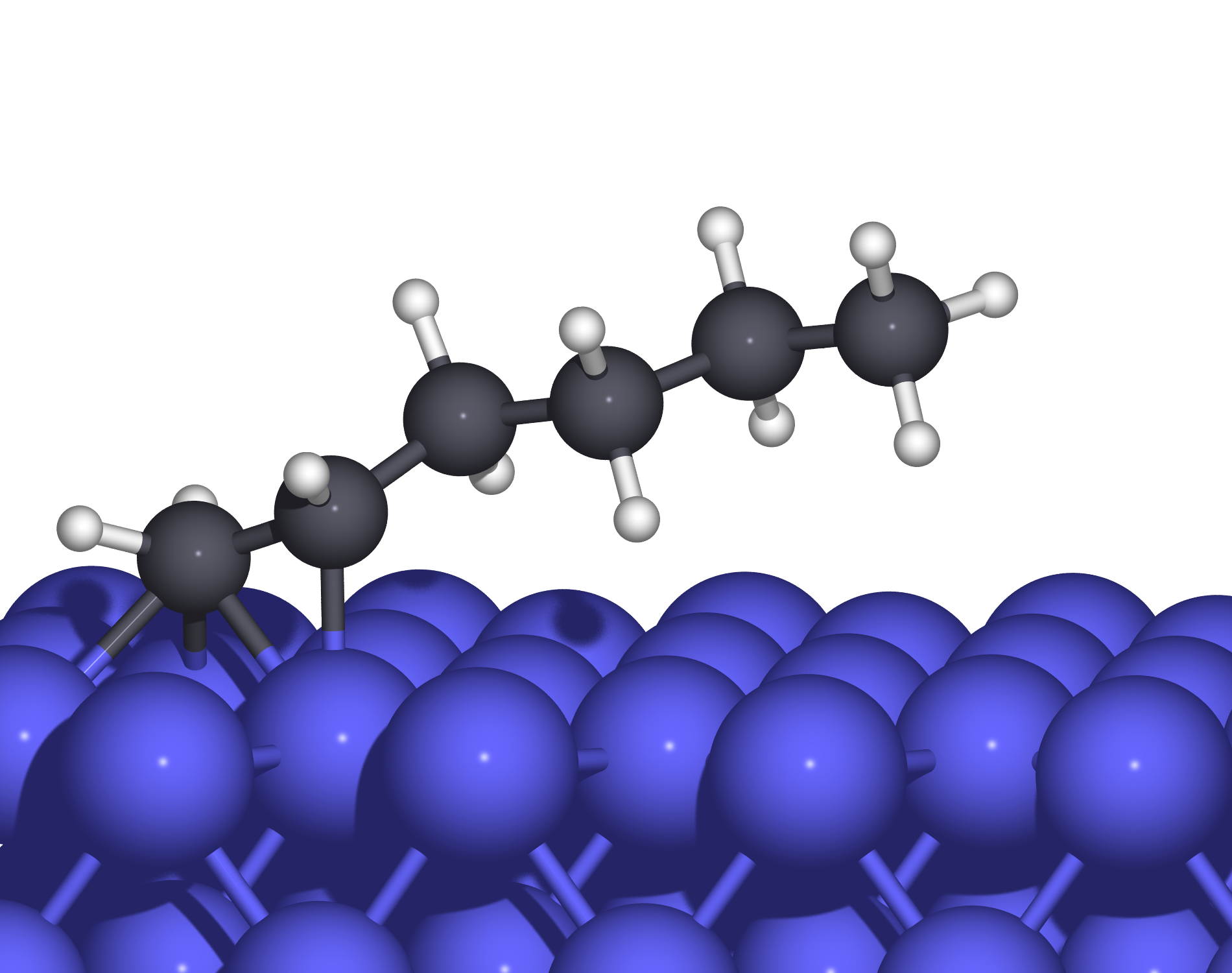}
        \caption{Lateral view.}\label{fig:Fe_finalrelax_lateral}
    \end{subfigure}
    \hfill
    \begin{subfigure}{0.23\textwidth}
        \centering
        \includegraphics[width=\textwidth]{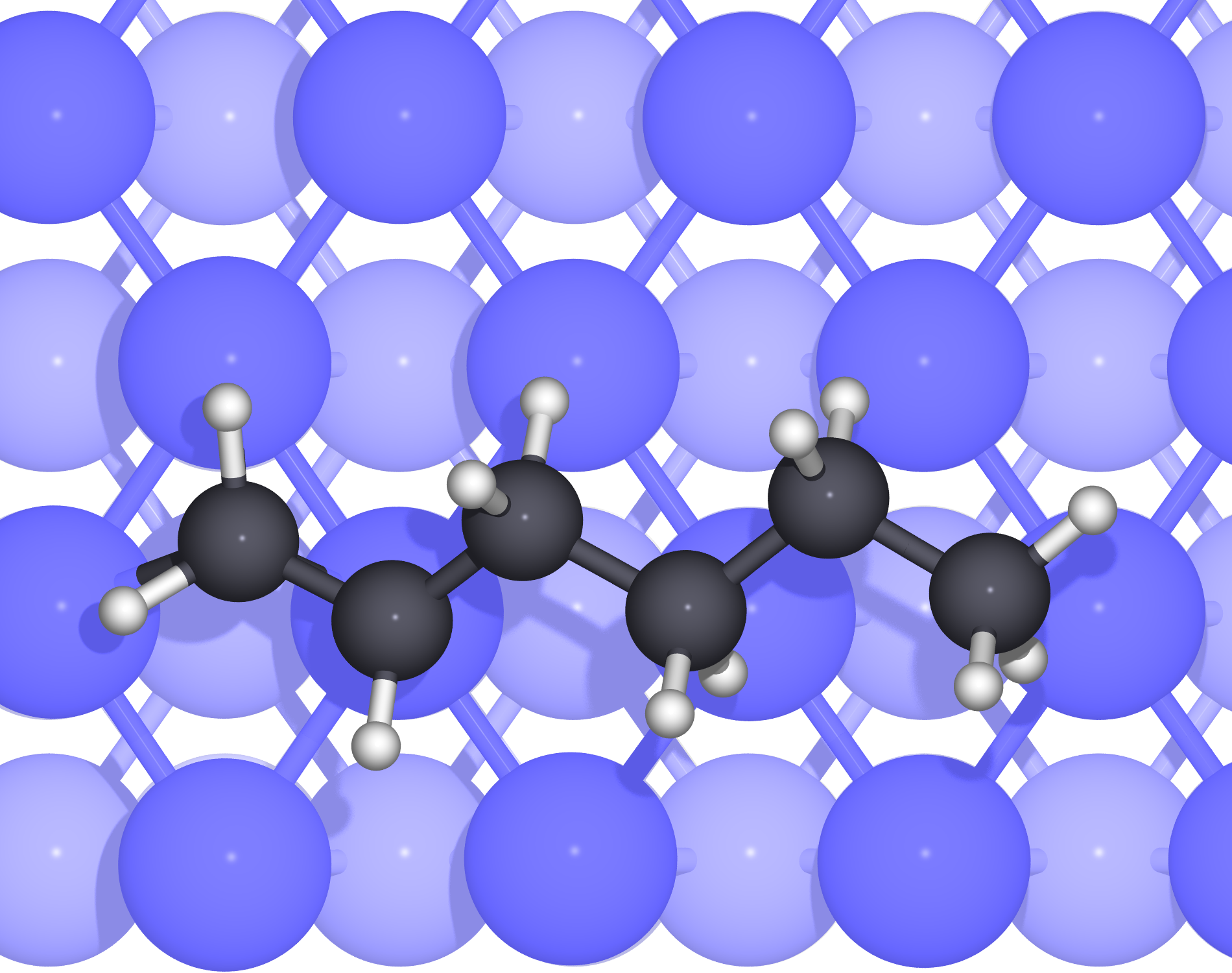}
        \caption{Top view.}\label{fig:Fe_finalrelax_top}
    \end{subfigure}
    \caption{Lateral (panel a) and top (panel b) representation of the adsorbed structure after the full geometrical optimization for the configuration with minimum energy (labelled 0). The different shades of colors for the Fe atoms represent the different depth of the atomic layers.}\label{fig:Fe_finalrelax}
\end{figure}

The last step in studying hexene adsorbed on Fe(110) is the full geometrical optimization of specific configurations. In particular, we selected the five configurations with lower energy from the screening. The final optimized adsorption energies are also reported in Tab.~\ref{tab:hex_Fe_table}, and the fully optimized structure corresponding to the global energy minimum is represented in Fig.~\ref{fig:Fe_finalrelax}.

This optimization procedure identified the most stable adsorption configurations for the hexene/Fe(110) system. In particular, the configuration labelled by 0 had the lowest adsorption energy, equal to $-1.83$ eV. This system showed the hexene double bonded C-C group anchored between the 3-fold and the on-top sites and the rest of the molecule slightly repelled from the Fe(110) surface. This hexene arrangement is the most favorable over Fe(110).

This preliminary screening method with partial structural optimization was in this case very effective in predicting the configuration associated to the energy minimum after the full optimization. Moreover, the ordering of the energies in the screening is almost identically reproduced after the optimization (with a single swap between configuration 8 and configuration 1). 

The computational cost of the screening was also quite contained, since for almost all configuration (with only three exceptions) the number of ionic steps in the screening was between 2 and 5, while the full optimizations of this system required between 70 and 120 ionic steps.


\section{Summary}

We presented \texttt{Xsorb}, a Python-based code that allows studying molecular adsorption on crystalline (reconstructed) surfaces. This program automatically generates several adsorption structures and screens their energy by DFT simulations. The most favorable adsorption geometry and energy is then identified. This approach can significantly help users, even non-expert, to study molecular adsorption, thanks to a user-friendly Python interface that can handle and simplify the identification of the most relevant adsorption configurations. We employed the \texttt{Xsorb} code to study 1-hexene adsorption on the Fe(110) surface. \texttt{Xsorb} can speed-up the study of adsorption processes in many research fields, such as catalysis, biomedicine, electrochemistry and tribology.



\section*{Acknowledgements}
These results are part of the SLIDE project that has received funding from the European Research Council (ERC) under the European Union’s Horizon 2020 research and innovation program. (Grant Agreement No. 865633).

 \bibliographystyle{elsarticle-num} 
 \bibliography{biblio}





\end{document}